# PORE ORDERING IN MESOPOROUS MATRICES INDUCED BY DIFFERENT DIRECTING AGENTS

Ana-Maria Putz [1], Savii Cecilia[1], Cătălin Ianăşi[1], Zoltán Dudás[1,2], Kinga Noémi Székely[2], Jiri Plocek[3], Paula Sfârloagă[4], Liviu Săcărescu[5], László Almásy*[2]

[1] Laboratory of Inorganic Chemistry, Institute of Chemistry Timisoara of Romanian Academy, Bv. Mihai Viteazul, No.24, RO-300223 Timisoara, Romania
[2] Wigner Research Centre for Physics, Institute for Solid State Physics and Optics, Hungarian Academy of Sciences, POB 49 Budapest-1525, Hungary
[3] Institute of Inorganic Chemistry, Academy of Sciences of the Czech Republic, Husinec-Rez 1001, 25068 Rez, Czech Republic
[4] R&D National Institute for Electrochemistry and Condensed Matter, P. Andronescu Street, No. 1, 300224 Timisoara, Romania
[5] Institutului de Chimie Macromoleculară „Petru Poni" Iaşi, Romania
*corr. author, Tel: +36-1-3922222

**Abstract**

Mesoporous silica particles of MCM-41 type were synthesized by sol-gel method from tetraethyl orthosilicate (TEOS) in 2-methoxyethanol and deionized water mixture in base conditions at room temperature. Ammonia or sodium hydroxides were used as catalysts and cetyl-trimethylammonium bromide (CTAB) and n-dodecyl-trimethylammonium bromide (DTAB) as structure directing agents. The porosities and the ordered structure have been analyzed using transmission and scanning electron microscopy, small angle neutron and X-ray diffraction, nitrogen adsorption, thermal analysis and FTIR spectroscopy. The samples consist of spherical particles of sub-micrometer size, with radially arranged pores. The comparison of the effect of the different surfactants and catalysts shows that by varying the surfactant type and their proportion, the pore sizes can be controlled. As compared to the commonly used ammonia catalyst, the use of NaOH as catalyst results in a much smaller porosity of the as-prepared materials. These materials are not resisting to the heat treatment at 700 ºC used for the template removal, and the ordered porous structure is completely lost.

1. **Introduction**

In recent years the progress in nanotechnology has motivated researchers to develop nanostructured materials for different applications, including biomedicine where the different types of nanoparticles serve as controlled delivery systems [1]. The porous biomaterials with large surface areas and large pore volumes are good candidates for drug delivery systems [2-4]. Drug carriers require biocompatibility and well defined morphological characteristics, and here the sol-gel methods of mesoporous silica materials production offer the simplicity of preparation and ease of control of their pore sizes adapted to the different drug molecules [5, 6].

Ordered mesoporous spherical silica particles can be obtained by using an ionic surfactant, such as cetyl-trimethylammonium bromide, CTAB. This was the first and even now is the most frequently used templating agent, for acid or base catalyzed sol-gel process or under hydrothermal conditions. Various pore morphologies can be achieved with different templating agents by adjusting their concentrations in the reaction mixture [7, 8].

Surfactants with shorter or longer alkyl chains can be used to control the pore size [9, 10]. Thus, silica particles synthesized with dodecyl-trimethylammonium bromide, DTAB, have a very small pore size, which allows a slowing down of the diffusion process and therefore minimizes the leakage in the system [11]. Modified Stöber synthesis can be used for preparation of micrometer sized silica particles with ordered pore structure [12-14]. The size of silica nanoparticles can be tuned by changing the synthesis conditions, the proportion of the solvents and reactants [15]. The increase of template agent chain length from 12 to 16 methyl groups results in increasing of the pore mean diameter [12, 16]. The two dimension parameters may also correlate: in an ammonia catalyzed synthesis, the particle size decreased as the chain length of the surfactant increased [17].

The pore sizes and the lattice periodicity depend primarily on the type of the surfactant [12, 18], but also organic co-solvents can affect the pore size acting as spacers inside the micelles [19]. Thus, ethoxyethanol acts not only as solvent but also as co-surfactant, controlling the morphology and pore structure. In alkali-catalyzed reactions, the co-surfactants promote TEOS water solubility facilitating the hydrolysis reactions [4, 20]. It was found that, compared with ethanol (the most used solvent in this kind of synthesis), methanol shows less inhibition effect on the development of long-range order, favouring the formation of silica spheres with hexagonally ordered mesopores [21].

2-Methoxyethanol belongs to the class of solvents known as glycol ethers which are notable for their ability to dissolve a variety of different types of chemical compounds and for their miscibility with water and other solvents. It is a porogen and confers hydrophobicity [22] 2-methoxyethanol was used as a co-solvent facilitating to dissolve PEG [23]. 2-Methoxyethanol could be used like a capping agent to help particles formation over a gel structure; the approach would be to passivate the silica nanoparticles (i.e., the particles in the sol) so that they cannot interconnect to form a gel. Also, a homogeneous gel state can be achieved by hydrolysis enhancement, respectively by lowering of solution temperature and also by addition of solvents like 2-methoxyethanol [24].

2-Methoxyethanol was previously used for the synthesis, by the sol-gel methods, of uniform spherical shape of silica nanoparticles [25]. 2-Methoxyethanol was also used as solvent and complexing agent in the synthesis of mixed silica–titania [26]. Moreover, the 2-Methoxyethanol allows, in one-step reaction, the preparation of silica-titania gels in the full composition range for very short gelation times. This is possible because of the use of 2-methoxyethanol, a protic polar solvent, which plays two different specific roles: it acts as a solvent as well as a stabilizer of titanium alkoxide towards the hydrolysis-precipitation reaction, by an accurate adjustment of the quantity of methoxyethanol in the mixture, controlling therefore the reactivity of the titanium precursor [27]. To our best knowledge, the use of 2-methoxyethanol as co-solvent in ordered porous silica synthesis has not been reported to date.

In the present study MCM-41 type materials have been prepared by using room temperature sol-gel hydrolytic route starting from tetraethyl orthosilicate (TEOS) as unique precursor, hydrolyzed in deionized water and 2-methoxyethanol mixture. Ammonia or sodium hydroxide was used as catalyst while CTAB or DTAB or their equimolecular mixture were used as templates. The effects of the two surfactants and of two catalysts on the sample morphologies have been studied.

## 2. Experimental

### 2.1 Sample synthesis

All chemicals were commercially available: tetraethyl orthosilicate (TEOS), (99%, for analysis, Fluka); 2-methoxyethanol (Sigma-Aldrich, 99.8%); $NH_4OH_{aq}$ (25%, SC Silal Trading SRL); NaOH (Chemopal); hexadecyl-trimethylammonium bromide (cetyl-trimethylammonium bromide) (Sigma-Aldrich); n-dodecyl-trimethylammonium bromide (Merck).

Silica samples with ordered porous structure we prepared by adapting earlier reported procedures [4, 14]. Szegedi and co-workers synthesized silica ordered matrix by using CTAB, TEOS in ethanol and water mixture and base catalyst $NH_3$ [14] Tan and co-workers synthesized porous silica nanoparticles by using CTAB, TEOS in 2-ethoxyethanol and water mixture and base catalyst $NH_3$ [4].

In the present work we used 2-methoxyethanol as co-solvent. The molar ratios of the reactants were TEOS / 2-methoxyethanol / $H_2O$ / $NH_3$(NaOH) / surfactant = 1:58:144:11(0.31):0.3 respectively. The base (catalysts) concentration was calculated as 5.25 mol/L, in order to keep constant the ionic strength in all reactants mixtures.

The synthesis procedure has been previously presented [28]. Six xerogel samples were prepared in sol-gel synthesis by using CTAB, DTAB or their equimolecular mixture, either in presence of ammonia or NaOH. Typically, CTAB (4.93 mg), or DTAB (4.17 mg), or their mixture, CTAB+DTAB (2.465+2.08 mg), were dissolved in a mixture of 117 mL of distilled water and 206.6 mL of 2-methoxyethanol. Subsequently 37.087 mL of ammonia (or 0.56 mg NaOH) were added. The mixtures were vigorously stirred in a closed vessel at room temperature for 30 min. Then, 10 mL of TEOS sol–gel precursor was slowly dripped in, and vigorous stirring was continued for 24 h. The resulted suspension was washed several times with distilled water in ultrasonic bath, and then centrifuged at 3500 rpm until the pH of the supernatant was close to neutral.

Ethanol was added to the separated precipitate and left for 24 h. After that, the material was dried at 40 °C for 3h, then heated up to 60 °C and dried during 9 h. These samples were labeled as 60. A part of the each dried sample has been further heated up to 500 °C, another part to 700 °C for 6 h, and the corresponding samples were labeled as 500 and 700 respectively (see Table 1).

Table 1. Sample names and synthesis conditions

| Sample name | Processing | Directing | Catalysts |
|---|---|---|---|

|  | Temperature (°C) | agent |  |
|---|---|---|---|
| CTAB-NH3-60 | 60 | CTAB | NH$_3$ |
| CTAB-NH3-500 | 500 | | |
| CTAB-NH3-700 | 700 | | |
| CTAB-NaOH-60 | 60 | | NaOH |
| CTAB-NaOH-500 | 500 | | |
| CTAB-NaOH-700 | 700 | | |
| DTAB-NH3-60 | 60 | DTAB | NH$_3$ |
| DTAB-NH3-500 | 500 | | |
| DTAB-NH3-700 | 700 | | |
| DTAB-NaOH-60 | 60 | | NaOH |
| DTAB-NaOH-500 | 500 | | |
| DTAB-NaOH-700 | 700 | | |
| Mix-NH3-60 | 60 | Mix | NH$_3$ |
| Mix-NH3-500 | 500 | | |
| Mix-NH3-700 | 700 | | |
| Mix-NaOH-60 | 60 | | NaOH |
| Mix-NaOH-500 | 500 | | |
| Mix-NaOH-700 | 700 | | |

## 2.2. Sample characterization

FTIR spectra were taken on KBr pellets with a JASCO –FT/IR-4200 apparatus. Textural properties of the composite silica samples were analyzed using nitrogen adsorption/desorption measurements at liquid nitrogen temperature (77K) with a Quantachrome Nova 1200e apparatus. Prior to the measurements the samples were degassed during 4 hours at room temperature. The specific surface area ($S_{BET}$) was calculated from the BET equation, the total pore volume ($V_p$) determined from the last point of adsorption and the pore diameter ($D_p$) determined by Barrett, Joyner and Halenda (BJH) method, from the desorption branches of the isotherms using a NovaWin software.

Small-angle neutron scattering (SANS) measurements were performed on the *Yellow Submarine* instrument at the BNC in Budapest (Hungary) [29]. Neutrons from a cold neutron source were monochromated by a mechanical velocity selector of 0.2 FWHM wavelength spread[30]. Sample-detector distances between 1.3m and 5.5m, and mean neutron wavelengths of 0.47 and 1.14 nm allowed to cover a q-range 0.6 – 4.2 nm$^{-1}$. The fine powder samples were filled in optical quartz glass cuvettes of 2 mm light path. Beam size at the sample was 8mm in diameter. Measurements were carried out at room temperature. The raw data have been corrected for attenuation, scattering contributions from the empty cuvette and the room background.

The morphological characterization of the samples was done by scanning and transmission electron microscopy (SEM and TEM). For SEM an INSPECT S (FEI Company, Olanda) instrument has been used. TEM investigations have been carried out with Hitachi High-Tech HT7700 Transmission Electron Microscope, operated in high contrast mode at 100 kV accelerating voltage. Samples have been prepared by drop casting from diluted dispersions of the nanoparticles in ethanol on 300 meshes holey carbon coated copper grids (Ted Pella) and vacuum dried.

X-ray powder diffraction (XRD) measurements were performed on Panalytical X'Pert Pro MPD diffractometer equipped with Cu-anode (Cu Kα; λ=0.15418 nm) and X'Celerator detector. Measurements were done at room temperature in transmission mode using Mylar foil over an angular range from 0.65 to 10° in 2θ.

The thermogravimetric analysis was carried out between 25 °C and 700 °C using a 851-LF 1100-Mettler Toledo apparatus in air flow and a heating rate of 5 °C min$^{-1}$.

## 3. Results and discussion

### 3.1. Thermal analysis

For all xerogel samples three stages of weight loss could be observed (Figs. 1,2). The first weight loss is accompanied by an endothermic effect on the DTA curves (Fig. 2), due to desorption of the physisorbed and chemisorbed water [31-33]. CTAB starts to decompose above 200 °C [34]. The surfactant transformations with temperature during calcination are well recognized up to ~250 °C. Trimethylamine is a product of the template degradation at lower temperatures. In the case of siliceous MCM-41 an intensive release of trimethylamine was observed within the temperature range 200–300 °C [31]. At a relatively low temperature 250–300 °C fragmentation of the alkyl chain takes also place producing shorter hydrocarbons [35, 36]. In our case the second weight loss accompanied by an endothermic effect was assigned to partial decomposition and reduction of organics to carbon. The last weight loss is accompanied by an exothermic effect on DTA, because of the low molecular weight fragments of the surfactant and carbonaceous material removal by oxidation to carbon oxides [37, 38]. The weight losses for all samples are presented in Table 2.

Table 2. Weight losses in the different temperature regions.

| Template | Catalyst | 25-180 | 180-375 | 375-700 | Total weight loss |
|---|---|---|---|---|---|
| CTAB |  | 11.62 | 42.61 | 7.70 | 53.18 |
| DTAB | NH$_3$ | 10.21 | 33.58 | 3.78 | 42.61 |
| Mix |  | 8.19 | 36.19 | 6.11 | 44.99 |
|  |  | 25-215 | 215-450 | 450-700 | Total weight loss |
| CTAB |  | 14.75 | 36.94 | 5.25 | 49.06 |
| DTAB | NaOH | 15.72 | 29.34 | 2.46 | 41.91 |
| Mix |  | 6.76 | 21.89 | 3.22 | 29.52 |

*3.1.1 The ammonia synthesized samples*

According to the TGA and DTA curves, the temperature domains for the weight losses were: 25-180 °C; 180-375 °C; 375-700 °C. As it was expected the highest total weight loss (53.18%) occurs in case of the samples synthesized with C$_{16}$TAB. It was observed that the total weight loss of the samples was increasing in the series: DTAB(C$_{12}$) < Mix < CTAB(C$_{16}$). The exothermic maxima of DTAB, Mix, CTAB ((figure 2a) were clearly distinct, at 324, 329 and 330 °C respectively [28].

In the case of samples synthesized with $NH_3$, two elimination steps of the surfactant were observed on DTG plot. Similar thermal behavior was observed by Denoyel et al. in synthesis of ZnO nanostructures using mix of CTAB and sodium dodecyl sulfate [39]. In the first step the surfactants started to degrade right after dehydration of water, followed by a second step when the residual surfactants decomposed until complete degradation. It can be assumed that, also in the present case, the two-step weight loss process can be due to decomposition and elimination first of the alkyl chain and then, at higher temperature, elimination of the head group. Finally, a rapid elimination by an auto combustion-like effect takes place (see the narrow TGA maximum in Fig. 2a).

*3.1.2 The sodium hydroxide synthesized samples*

For the samples synthesized with sodium hydroxide the weight loss temperature domains shifted to higher temperatures: 25-215 °C; 215-450 °C; 450-700 °C (compared to the ammonia synthesized samples). The total weight loss of the samples increases in the series: $CTAB(C_{16})$ < $DTAB(C_{12})$ <Mix . The exothermal maxima for CTAB, DTAB, Mix, (Fig. 2a) were distinctly noted at 337, 343 and 355 °C respectively [28]. For each sample case the total weight losses have lower values (Table II), compared to the samples synthesized with ammonia.

*3.1.3. Comparison of the $NH_3$ and NaOH synthesized samples*

The thermal analysis results show that the samples synthesized with NaOH are more stable according to the broadening of the second weight loss domain which starts at somewhat higher temperature than for samples with ammonia. It can be attributed to the higher reactivity of NaOH, with both, the templating agent and the matrix. In the case of samples synthesized with NaOH, an overlap of the two elimination steps is observed, so there is a partial overlapping in the decomposition and elimination of the alkyl chains and the core structure.

The carbon elimination by oxidation is slower in the NaOH samples than in the $NH_3$ ones, probably due to the reduced porosity hindering of the surfactant in the matrix. In general, disordered structures, like mesoporous silica material prepared using polymerization of silicate anions surrounding surfactant micelles in the presence of organic salts have higher thermal stability than the ordered ones, such as MCM-41 [40].

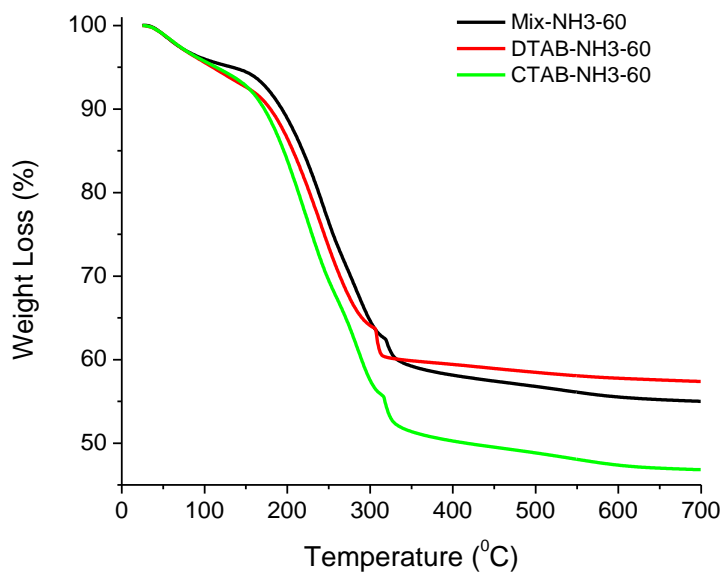

a

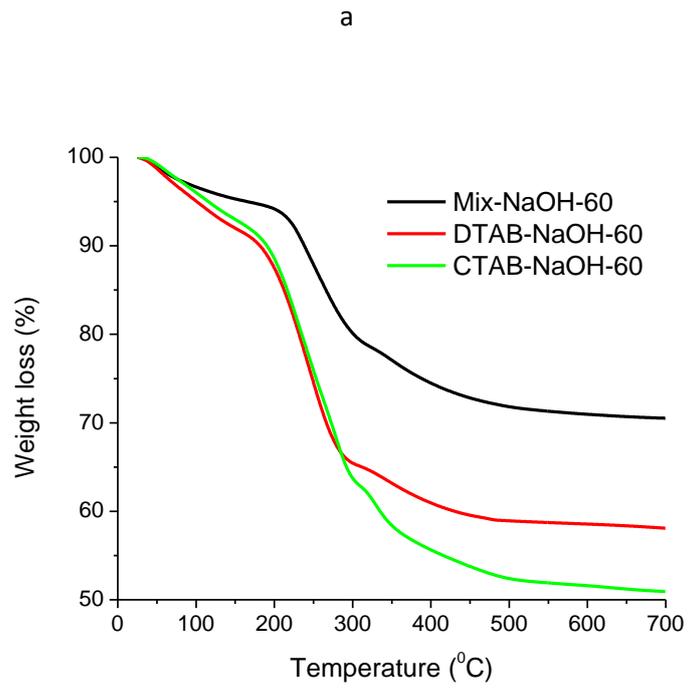

b

Figure 1. *TGA curves for the samples obtained with NH$_3$ (a) and NaOH (b) catalysis*

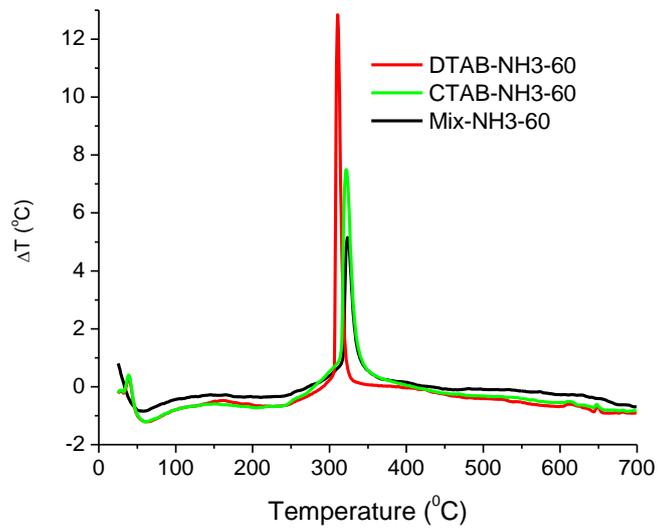

a

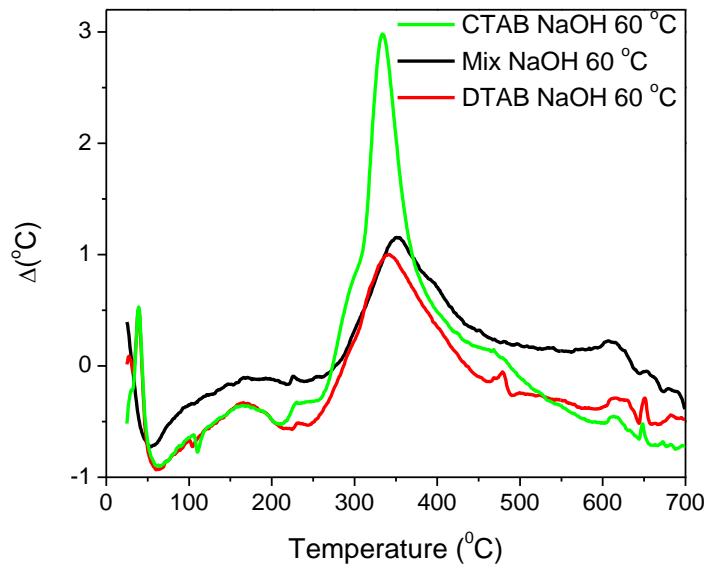

b

Figure 2. *DTA curves for the samples obtained with $NH_3$ (a) and NaOH (b) catalysis*

### 3.2 FT-IR Spectra

FT-IR spectra of the xerogels dried at 60 ºC ($NH_3$ catalyst) are presented in Figure 3. No obvious differences could be observed between IR spectra of the ammonia and sodium hydroxide synthesized samples.

As expected, all samples (xerogels and thermally treated materials) show the main specific vibration bands for the silica skeleton at 1050, 800 and 450 $cm^{-1}$. They can be assigned to the asymmetric stretching [41, 42], symmetric stretching [42, 43] and its bending rocking mode vibration of the Si-O-Si network [42, 44-46], respectively.

For all the thermally treated samples the position of asymmetric stretching vibration band is shifted to higher values by approximately 20 $cm^{-1}$, which is related to the densification of the silica network [47].

The presence of the silanol groups were confirmed by the presence of the band centered about 960 $cm^{-1}$, which is associated with the stretching mode of the Si-OH groups [45].

The 1640 $cm^{-1}$ vibration band belongs to the molecular water [45]. The broad band centered at around 3470–3450 $cm^{-1}$ corresponds to the overlapping of the O-H stretching bands of hydrogen-bonded water molecules (H-O-H...H) and SiO-H stretching of surface silanols hydrogen-bonded to molecular water (-SiOH...$H_2O$) [43]. This can be attributed to the higher hygroscopicity of the calcined samples, due to the fact that in the xerogels the template molecules fill the pores, but after the calcination the pores are freed and the water molecules have more space to occupy.

The third group of characteristic bands is due to the surfactant molecules. In the non-calcined samples the bands at 2920 and 2850 $cm^{-1}$ (see Fig. 4) are the asymmetric and symmetric stretching vibration) groups of the methylene chain, respectively [48-50]. The band around 1480 cm$^{-1}$, is ascribed to C–H bond stretching vibration of the alkyl chains [6, 51]. These bands almost completely disappear in the calcined samples.

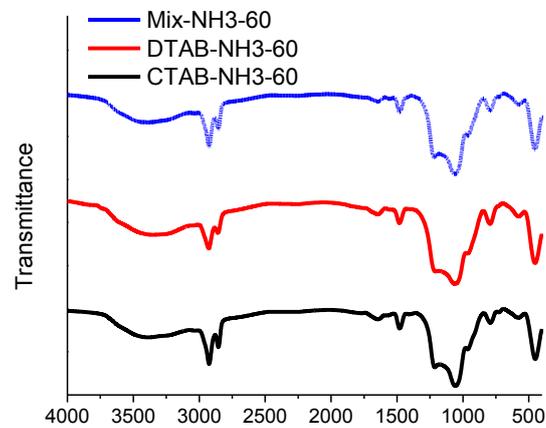

Figure 3. *FT-IR spectra of the silica xerogels made with NH$_3$ catalyst*

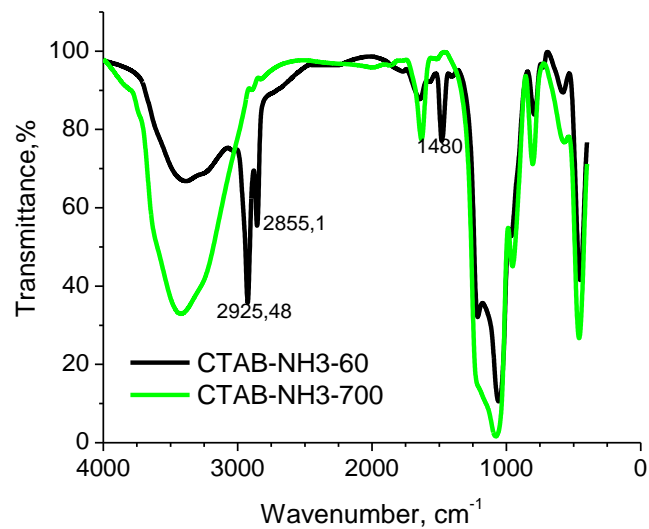

Figure 4. *FT-IR spectra of the CTAB-NH3 sample before and after calcination at 700ºC*

### 3.3. Nitrogen adsorption
*3.3.1. Ammonia synthesized samples*
In figure 5 are presented the nitrogen adsorption/desorption isotherms of the samples prepared with the NH$_3$ catalysts. The xerogel samples have IV type isotherm with H3 characteristic hysteresis loops attributable to the materials with the slit shaped pores [52, 53]. After templating agent removal at 500 ºC all isotherms are of IV type, characteristic for MCM-41 materials with elongated pores [54,55]. At 700 ºC the DTAB and Mix samples retain the IV type isotherm, while the CTAB-NH3-700 exhibit a hysteresis shape H4, indicating partial degradation of the pores. Very similar isotherm has been shown in base catalyzed MCM-41 materials after partial disintegration at high temperature [55].

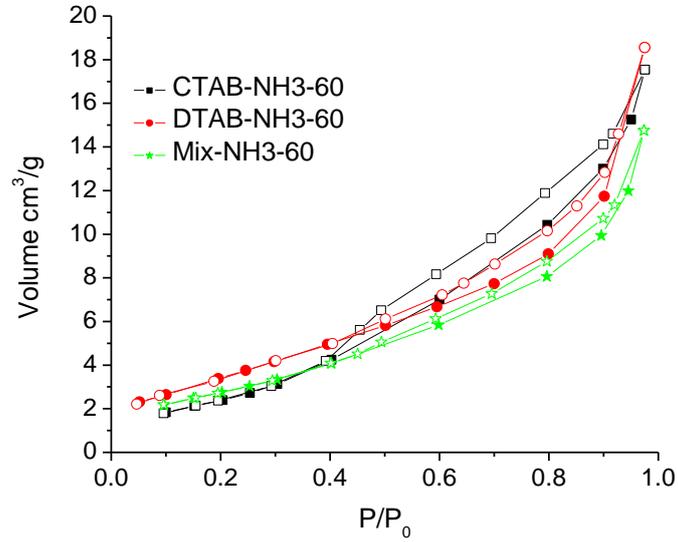

a

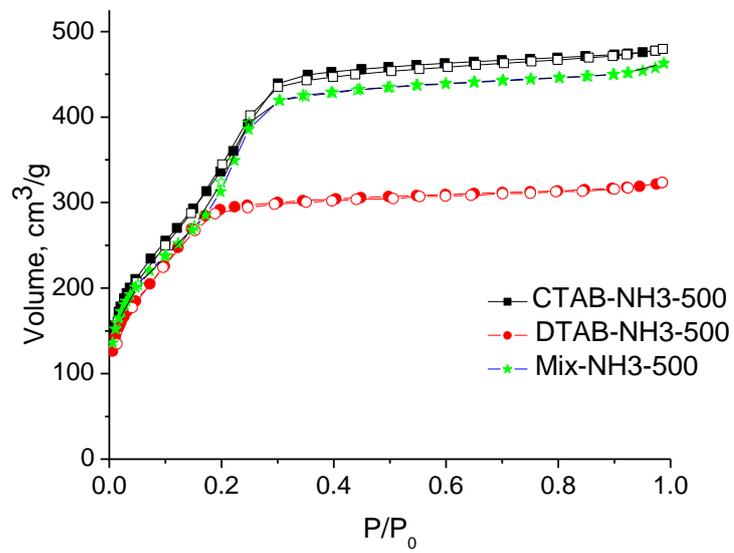

b

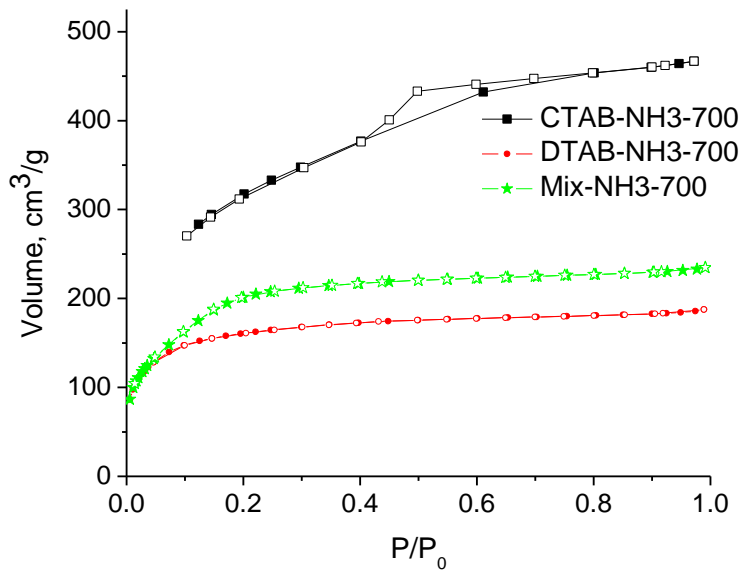

c

Figure 5. *Nitrogen adsorption/desorption isotherms of CTAB-NH3-60 (squares), DTAB-NH3-60 (circles), Mix-NH3-60 (stars), (a), CTAB-NH3-500, DTAB-NH3-500, Mix-NH3-500 samples (b), and CTAB-NH3-700, DTAB-NH3-700, Mix-NH3-700 (c).*

For the xerogel samples the average pore sizes increases in the following order: DTAB < mix < CTAB (Table 3). All samples possess porosity in the microporous region also; the average pore size is 2 - 2.3 nm. In the calcined samples the average pore size is lower compared to the corresponding xerogel samples. Total pore volume evolution with temperature is indicating the densification of the sodium hydroxide derived silica network. In the case of ammonia catalyzed samples, the pore volume increased by 10-25 times when the xerogels were heated up to 700 °C. The consolidation of the ordered network by thermal treatment was also proved by the FT-IR, SANS and XRD measurements.

Table 3. The textural properties of the synthesized samples

| Sample name | $D_p$ (nm) | $S_{BET}$ (m²/g) | $V_p$ (cm³/g) |
|---|---|---|---|
| CTAB-NH3-60 | 3.63 | 10.5 | 0.027 |
| CTAB-NH3-500 | 3.18 | 1431.35 | 0.743 |
| CTAB-NH3-700 | 2.02 | 1060 | 0.722 |
| DTAB-NH3-60 | 2.77 | 13.4 | 0.029 |
| DTAB-NH3-500 | 2.58 | 990.19 | 0.501 |
| DTAB-NH3-700 | 2.02 | 504 | 0.295 |
| Mix-NH3-60 | 3.17 | 10.8 | 0.023 |
| Mix-NH3-500 | 3.18 | 1383.49 | 0.718 |
| Mix-NH3-700 | 2.31 | 903 | 0.472 |
| CTAB-NaOH-60 | 4.54 | 8.93 | 0.016 |
| CTAB-NaOH-500 | 2.31 | 918.57 | 0.468 |
| CTAB-NaOH-700 | 2.77 | 6.77 | 0.009 |
| DTAB-NaOH-60 | 4.54 | 6.05 | 0.011 |
| DTAB-NaOH-500 | 2.31 | 517.50 | 0.267 |
| DTAB-NaOH-700 | 2.65 | 6.42 | 0.008 |
| Mix-NaOH-60 | 4.54 | 6.20 | 0.017 |
| Mix-NaOH-500 | 2.65 | 28.48 | 0.025 |
| Mix-NaOH-700 | 2.77 | 6.66 | 0.009 |

Specific surface area of xerogel samples was of ~ 10-13 $m^2/g$, probably due to hindered surface of the pores by confined surfactants. The thermally treated samples showed largely increased pore volume and specific surfaces. The maximum specific surface area (1431 $m^2/g$) was obtained for the sample obtained with CTAB and $NH_3$ and calcined at 500 ºC. On further heating it decreased to 1060 $m^2/g$. These values are consistent with the typical values reported for ordered porous materials prepared with CTAB [9, 14]. In a previous study, BET surface areas were reported, as 1157 and 1099 $m^2/g$ for MCM-41 with CTAB and with DTAB, with a total volume of pores of 0.98 and 0.84 $cm^3/g$, respectively [9]. In our synthesis, the DTAB-NH3-700 sample had somewhat smaller surface areas, 990 and 504 $m^2/g$ after calcination at 500 and 700 ºC, respectively.

When mixing the two surfactants, the characteristics porosity values are in between those of the single surfactant templated samples. Actually they are closer to those of the samples prepared with CTAB. Hence, it can be assumed that for fine tuning the porosity by using mixed surfactant template, one should balance the shorter chain surfactant and the longer chain surfactant fractions.

### 3.3.2. Sodium hydroxide synthesized samples

In Fig. 6 the nitrogen adsorption/desorption isotherms are shown for the xerogels and calcined samples obtained in the NaOH catalyzed synthesis. Both the dried and the 700 °C treated samples showed type II isotherms, characteristic for nonporous materials, though their structures were different: the dried samples were nonporous because of the pore filling by the surfactants, while in the calcined samples the pores were collapsed during the high temperature thermal treatment. The DTAB and CTAB templated samples calcined at 500 °C show IV type isotherm with specific surfaces 518 and 918 $m^2/kg$, respectively, while the sample prepared with mixed surfactants collapsed, showing II type isotherm.

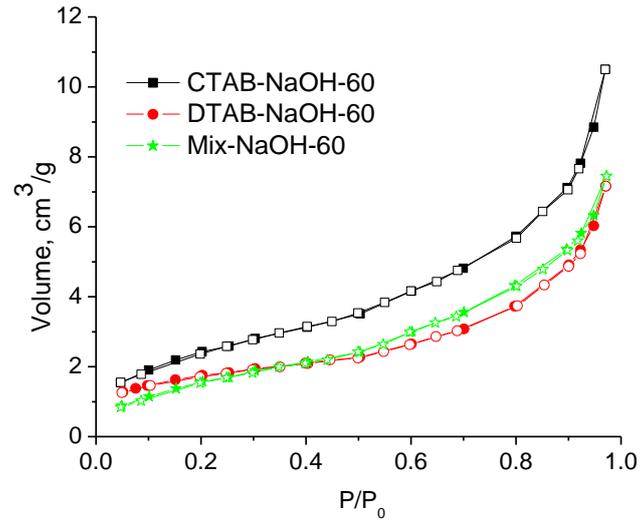

a

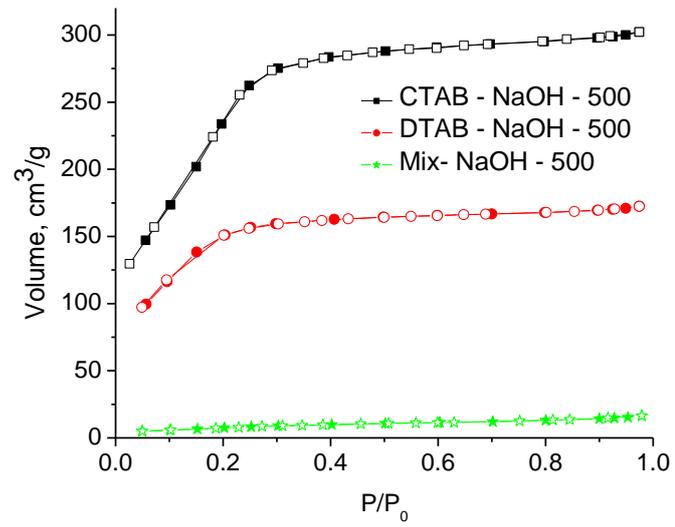

b

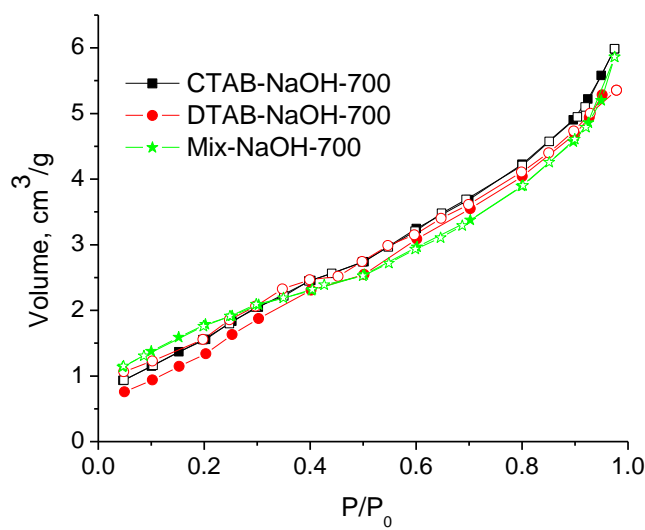

c

Figure 6. *Nitrogen adsorption/desorption isotherms of samples CTAB-NaOH-60, (squares) DTAB-NaOH-60(circles), Mix-NaOH-60(stars) (a); CTAB-NaOH-500, DTAB-NaOH-500, Mix-NaOH-500 (b) and CTAB-NaOH-700, DTAB-NaOH-700, Mix-NaOH-700 (c).*

### 3.4. Electron microscopy

Scanning and transmission electron microscopy reveal the shape and inner structure of the synthesized particles. Two representative SEM images of the CTAB templated samples are shown in Fig 7. All samples show spherical particles with sizes 0.5-0.8 μm.

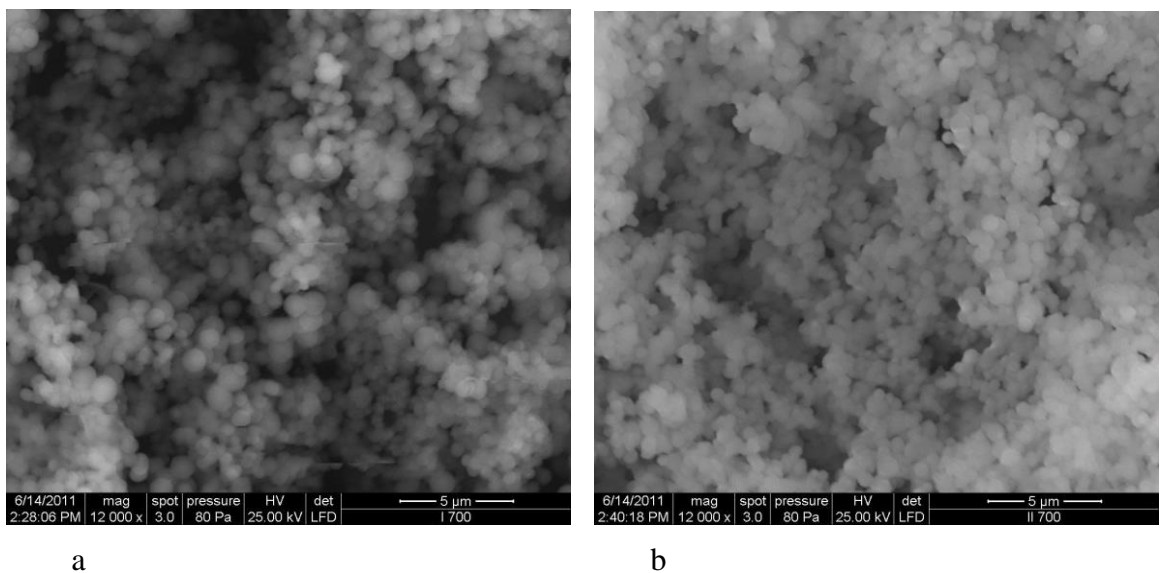

a                       b

Figure 7. *SEM micrographs of samples CTAB-NH3-700 (a) and CTAB-NaOH-700(b)*

The TEM images of the calcined CTAB-NH3-700 samples are shown in Figure 8. Particles are perfectly spherically shaped, their diameters vary between 200 and 800 nm. Higher magnification reveals the ordered nanometer size channels, dominantly arranged in radial directions. The channel distance as calculated from the images was 4.1 nm for CTAB-NH3-60 samples and 3.2 nm for CTAB-NH3-700.

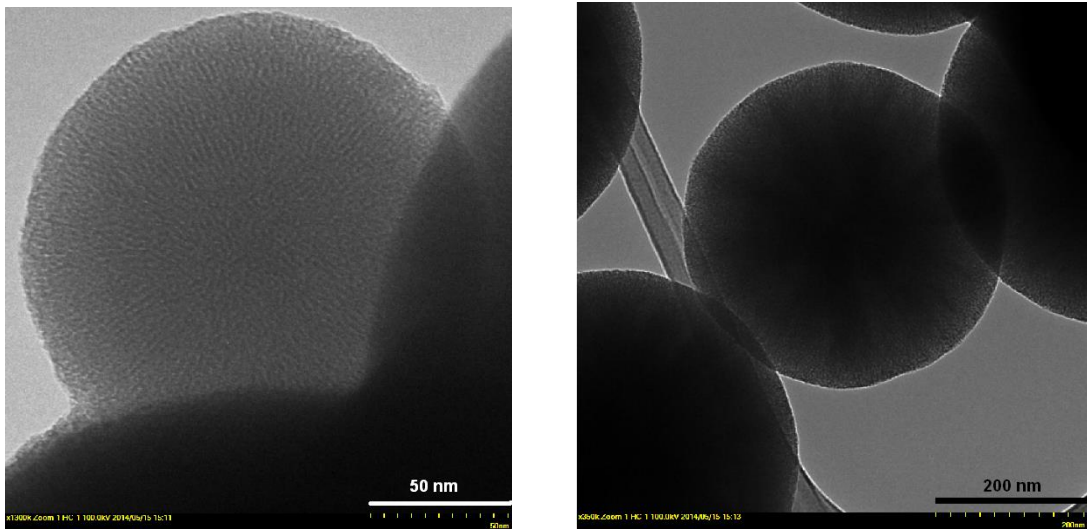

Figure 8. *TEM images of silica particles for samples CTAB-NH3-700*

In contrast to the ammonia synthesized samples, the CTAB-NaOH-700 submicron-spheres are homogeneous at the studied magnification; TEM does not reveal the characteristic long channel porosity in these particles (Fig. 9). Particles with porous structure are very rare, practically absent on the studied parts of TEM grids. All of them are highly dense and have smooth surfaces.

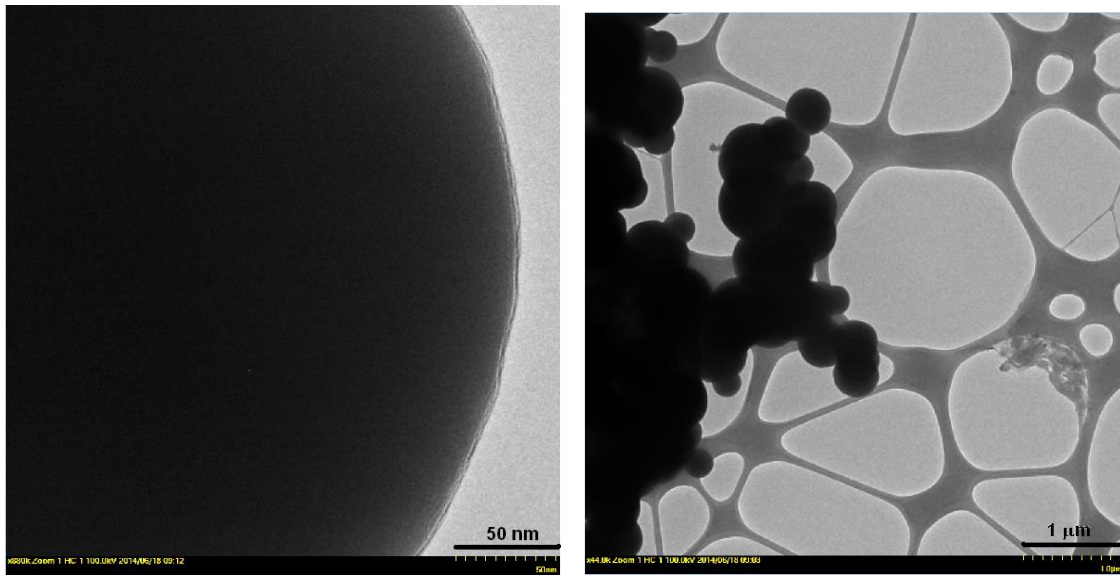

Figure 9. *TEM images of silica particles for samples CTAB-NaOH-700*

### 3.5 Small angle neutron scattering (SANS) and X-ray diffraction

Small-angle scattering allows to study the structure of the materials at length scales 1-100 nm, providing a volume-average of the inner structure of the materials. SANS scattering curves of the dry silica gels treated at 60 °C and 700 °C are shown in Figures 10,11,12.

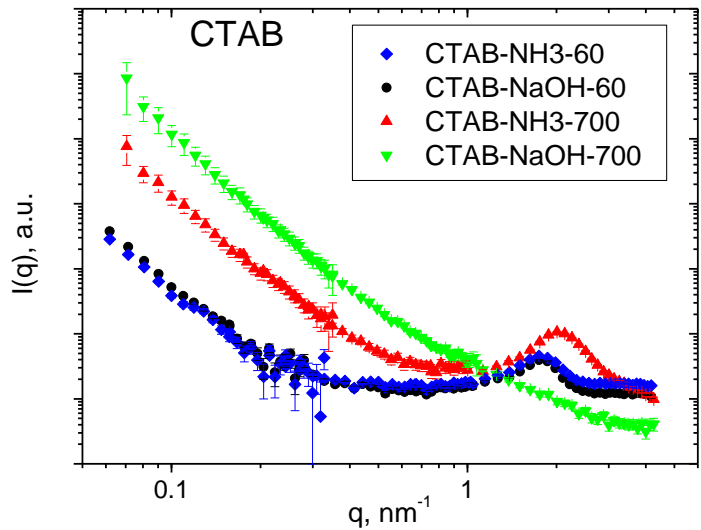

Figure 10 *SANS data of samples prepared with CTAB*

In Fig. 10 the scattering curves of samples made with CTAB are compared. The curves of the 60°C dried materials show the typical structure of the ordered porous silica gel – an interference peak at $q_0 = 1.77$ nm$^{-1}$, corresponding to $2\pi/q_0 = 3.55$ nm interplanar distance. After calcination at 700°C, the interference peak disappears for the CTAB-NaOH sample, while it shifts to $q_0 = 2.05$ nm$^{-1}$ for the other sample corresponding to a decrease of the characteristic distance to 3.06 nm.

The silica gels prepared using the DTAB templating agent, show the same structural features, see Fig. 11. For both non-calcined samples the peak position at $q_0 = 1.94$ nm$^{-1}$ corresponds to repeat distance of 3.24 nm, which decreases to 2.74 nm ($q_1 = 2.29$) after calcination for the NH$_3$ catalyzed sample, and the ordered structure completely disappears for the NaOH sample.

For the samples prepared with the mixture of the surfactants, the situation is overall similar, and the peak positions are 1.72 and 2.01 nm$^{-1}$, before and after calcination (Fig. 12).

For the calcined samples, the scattering at small angles follows the $I(q) \sim q^{-4}$ behavior, which displays in the log-log representation as a straight line with slope -4, as can be seen in Fig. 11. It corresponds to Porod's law, and shows that surface of the silica microspheres is smooth [56]. For the non-calcined samples, the slopes are slightly lower, about -3.8, which can be attributed to their less smooth surface [57]. More precise measurements extending to lower q values would be necessary to resolve these differences.

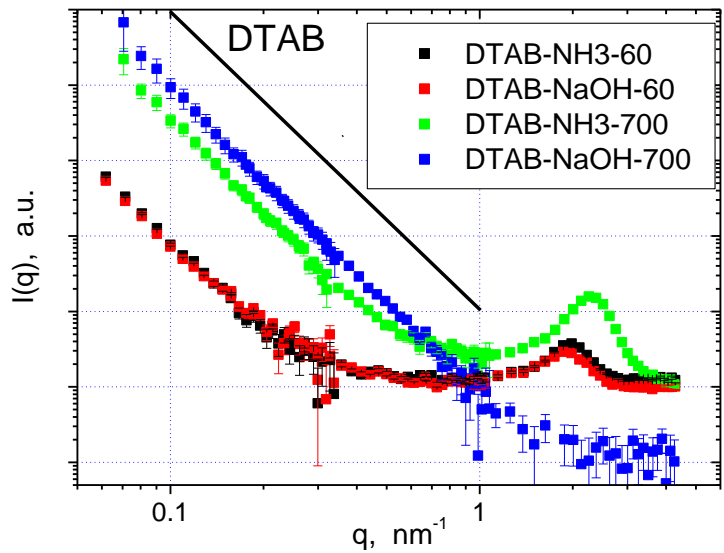

Figure 11. *SANS data on samples made with DTAB. The solid line is an asymptote of the power law $I(q) \sim q^{-4}$*.

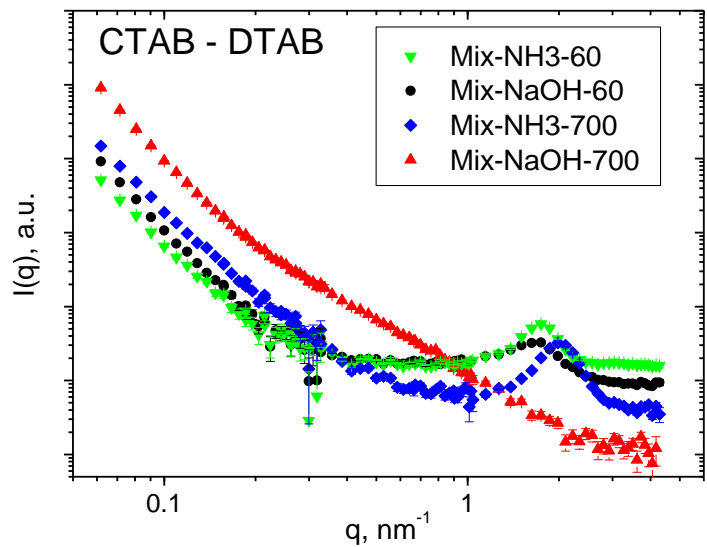

Figure 12. *SANS data on samples made with CTAB-DTAB mixed templating agent*

X-ray diffraction was performed for the samples prepared with the two templating agents and the two catalysts, before and after calcination at 700°C (Fig. 13). The data reveal the same structural development as observed in SANS. The better angular resolution of XRD as compared to SANS allows detecting a very week and broad higher order peak for most of the samples. In MCM-41 type materials with well developed hexagonal structure of cylindrical pores, usually 3 to 5 reflections are visible, depending on the extent and quality of the long range order [9, 14, 40]. For the present samples only the weak second order peak can be discerned, indicating that the structure is not so perfect and does not show long range periodicity. In the NaOH synthesized samples, both the first and the second order peaks disappear after calcination.

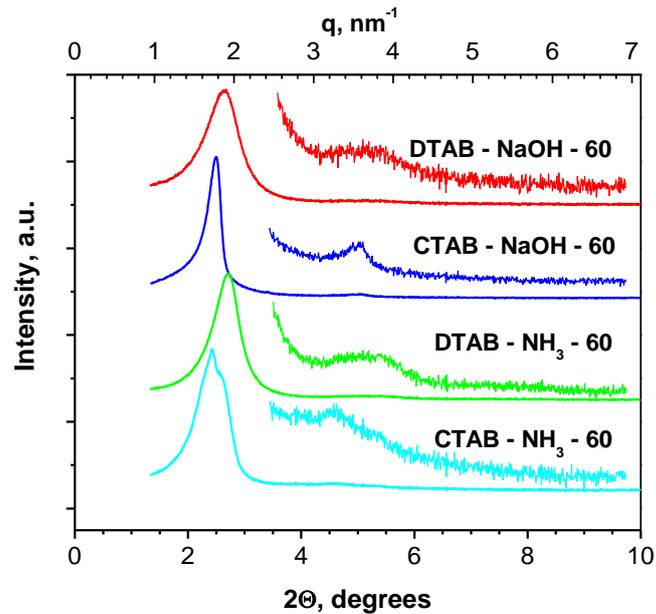

a

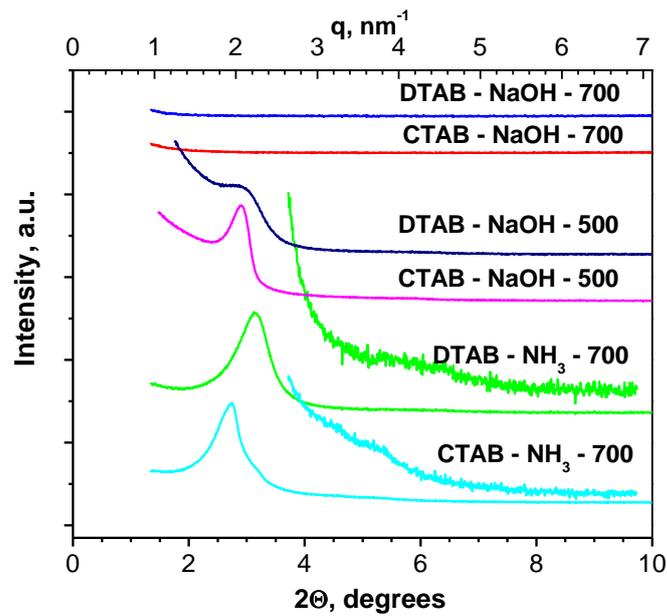

b

Figure 13. *Powder X-ray diffraction data of samples heated to 60 °C (a) and calcined at 500 and 700 °C (b). The higher angle parts of the scattering curves are multiplied by a factor of 10.*

Both diffraction experiments indicate that after calcination, the removal of the surfactant causes shrinkage of the distance between the pores. This shrinkage amounts in average to 15% for the samples prepared with ammonia. In the samples prepared with NaOH, the interference peak completely disappears, indicating that the ordered porous structure is collapsed, and no characteristic interpore distance remains. Hence, those samples were not resistant to the calcination at 700 ºC.

Although the use of long alkyl chain surfactants as templating agents leads usually to hexagonally arranged cylindrical pores, different structures may also form. Thus, in recent study of CTAB templated spherical silica particles, the electron microscopy revealed a complex core-shell structure, in which the core of the particle had apparently cubic structure, while the outer shell contained radially oriented nearly parallel channels [14]. Similar radial channels were also observed in some of our particles (Fig. 8a). Pauwels et al [58] compared spherical and ordinary MCM-41 materials by high resolution TEM, and showed that in the spherical particles only short range hexagonal ordering exists, but it does not extend to the whole particle volume. As a result, an ill-ordered porous structure emerges, with an average spherical symmetry.

## 4. Conclusions

Spherical silica mesoporous particles of MCM-41 type have been synthesized by sol-gel method in base conditions using two templating agents, surfactants CTAB and DTAB. 2-Methoxyethanol was used as co-solvent, and the effects of two different bases used as catalyst, sodium hydroxide and ammonia have been compared.

Both templating agents and both catalysts lead to submicron sized spherical particles with elongated parallel pores. The pore sizes and the interpore distances were higher for the CTAB templated samples, in accordance with the longer alkyl chain of this surfactant. The thermogravimetric and differential thermal analysis indicated that the xerogels, catalyzed by sodium hydroxide were more stable than those prepared with ammonia. For the samples synthesized with sodium hydroxide the burning of the template surfactant takes place at higher temperatures, compared to the samples made with $NH_3$. On the other hand the porous network in samples prepared with $NH_3$ is more resistant to calcination: the ordered porous structure disintegrates in the NaOH catalyzed samples below 700 °C, while in the $NH_3$ catalyzed samples the ordered pore structure remains, accompanied with shrinkage of the interpore distance.

Nitrogen adsorption showed low specific surfaces in the template-filled samples, while after calcination to 500 °C the specific surfaces in the $NH_3$ samples increased to the level of 1000 $m^2/g$, typical for MCM-41 type porous materials. Further heating to 700 °C resulted in pore collapsing in the NaOH catalyzed materials, while the porosity in the ammonia catalyzed samples diminished by about 30%.

When using mixture of two surfactants, the porosity became closer to the porosity of the CTAB templated samples, indicating that in case of an equimolar mixture of surfactants, the properties of the longer surfactant dominate the pore structure of the resulting material. The pore size and inter-pore distances can be varied changing the ratio of the templating surfactants.

**Acknowledgements**

Authors thank the Romanian Academy and the Inter-Academic Exchange Program between Academy of Sciences of the Czech Republic and Romanian Academy. This work was supported by project KMR12_1_2012_0226 granted by the National Development Agency (NDA) of Hungary.